# Lead-free Hybrid Perovskite: An Efficient Room Temperature Spin Generator via Large Interfacial Rashba effect

Lei Han,[1,‡] Qian Wang,[1,‡] Ying Lu,[2,3,‡,*] Sheng Tao,[4] Wenxuan Zhu,[1] Xiaoyu Feng,[5] Shixuan Liang,[1] Hua Bai,[1] Chong Chen,[1] Kai Wang,[4] Zhou Yang,[3] Xiaolong Fan,[5] Cheng Song,[1,*] Feng Pan[1,*]

[1]Key Laboratory of Advanced Materials (MOE), School of Materials Science and Engineering, Tsinghua University, Beijing 100084, China

[2]Technological Institute of Materials & Energy Science (TIMES), Xi'an Key Laboratory of Advanced Photo-Electronics Materials and Energy Conversion Device, School of Electronic Information, Xijing University, Xi'an 710025, China

[3]Department of Materials Physics and Chemistry, School of Materials Science and Engineering, University of Science and Technology Beijing, Beijing 100083, China

[4]Institute of Optoelectronics Technology, School of Physical Science and Engineering, Key Laboratory of Luminescence and Optical Information, Ministry of Education, Beijing Jiaotong University, Beijing 100044, China

[5]The Key Lab for Magnetism and Magnetic Materials of Ministry of Education, Lanzhou University, Lanzhou 730000, China

ABSTRACT: Two-dimensional (2D) hybrid organic-inorganic perovskite (HOIP) demonstates great potential for developing flexible and wearable spintronic devices, by serving as spin sources *via* the bulk Rashba effect (BRE). However, the practical application of BRE in 2D HOIP faces huge challenges, particularly due to the toxicity of lead, which is crucial for achieving large spin-orbit coupling, and the restrictions in 2D HOIP candidates to meet specific symmetry-breaking requirements. To overcome these obstacles, we design a strategy to exploit the interfacial Rashba effect (IRE) of lead-free 2D HOIP $(C_6H_5CH_2CH_2NH_3)_2CuCl_4$ (PEA-CuCl), manifesting as an efficient spin generator at room temperature. IRE of PEA-CuCl originates from the large orbital hybridization at the interface between PEA-CuCl and adjacent ferromagnetic layers. Spin-torque ferromagnetic resonance measurements further quantify a large Rashba effective field of 14.04 Oe per $10^{11}$ A m$^{-2}$, surpassing those of lead-based HOIP and traditional all-inorganic heterojunctions with noble metals. Our lead-free 2D HOIP PEA-CuCl, which harnesses large IRE for spin generation, is efficient, nontoxic, and economic, offering huge promise for future flexible and wearable spintronic devices.

INTRODUCTION: Efficient spin generation is crucial for low-power spintronic devices, represented by magnetic random-access memory and spin transistors.[1-3] Traditionally, inorganic single layers or heterojunctions have been employed as spin generators, based on the well-established mechanisms like the spin Hall effect[4] or the Rashba effect.[5, 6] In fact, organic materials offer distinct advantages over inorganic counterparts for spintronic applications, including flexibility, versatility, and longer spin coherence time, which are particularly beneficial for wearable technologies.[7-11]

Beyond conventional organic semiconductors (*e.g.*, Alq$_3$,[12] conjugated polymers,[7] V(TCNE)$_x$[9]), two-dimensional (2D) hybrid organic-inorganic perovskite (HOIP) present several benefits.[13-26] These materials feature highly tunable electronic structures and convenient synthetic methods, making them promising candidates for electrically controllable and cost-effective spintronic devices. Hence, as a fundamental effect for efficient spin generation, the bulk Rashba effect (BRE) has been extensively investigated in 2D HOIP[13-17, 19, 20, 24-27] using various techniques such as nonlinear and ultrafast spectroscopies,[15, 25, 26] the circular photogalvanic effect,[17, 19] angle-resolved photoemission spectroscopy,[14] *etc.*[13, 16, 20, 24]. These studies focus on identifying the characteristic Rashba spin-splitting bands, which exhibit horizontal splitting along the momentum $k$ direction with spin-momentum locking,[5] enabling the conversion of a longitudinal current into a transverse spin current.

Despite these advancements, several critical challenges remain in utilizing 2D HOIP as efficient spin generators *via* the BRE. Lead-based 2D HOIP [such as (PEA)$_2$PbI$_4$[15] and (*R/S*-NEA)$_2$PbI$_4$[21]] are necessary to produce large spin-orbit coupling (SOC) for a large BRE. However, the toxicity of lead poses serious health risks, limiting their widely application. Moreover, for BRE to occur, the inversion symmetry must be broken, which either necessitates the introduction of chiral molecules[21] or the selection of specific 2D HOIP with symmetry-breaking structures,[18] thereby limiting the range of suitable materials. Furthermore, the semiconducting or insulating nature of 2D HOIP limits the current density available for conversion into spin, which hinders efficient spin current generation *via* the BRE.

To address these challenges, we propose a strategy to exploit the interfacial Rashba effect (IRE) of lead-free 2D HOIP. First, the interface between 2D HOIP and other materials with large orbital hybridization can be expected to produce large IRE,[5, 28, 29] which eliminates the need for lead to induce large SOC as required in BRE. Second, at these interfaces, the inversion symmetry is naturally broken, expanding material candidates. Third, the IRE can take use of the high conductivity of the material adjacent to the semiconducting 2D HOIP for efficient spin generation, a principle similar to that used in traditional insulator/ferromagnet inorganic systems.[30-32] As a result, employing IRE in lead-free 2D HOIP is promising for spin generation, but remains unexplored.

In this work, we demonstrate that the lead-free 2D HOIP $(C_6H_5CH_2CH_2NH_3)_2CuCl_4$ (PEA-CuCl) serves as an efficient room temperature spin generator *via* large IRE. We selected PEA-CuCl for its nontoxicity and well-characterized properties,[33-35] while maintaining a similar 2D structure to lead-based perovskites (such as $(PEA)_2PbI_4$[15]), allowing for direct comparison. We first theoretically predict a substantial IRE at the interface between PEA-CuCl and ferromagnetic metal Ni, supported by calculations of the characteristic Rashba spin-splitting electronic structure. This large IRE comes from strong interfacial orbital hybridization. Subsequently, experimental determination of the IRE-induced Rashba effective field is achieved by systematic spin-torque ferromagnetic resonance measurements, yielding 14.04 Oe per $10^{11}$ A m$^{-2}$. This value surpasses those observed in lead-based 2D HOIP and other traditional all-inorganic heterojunctions, demonstrating the potential for lead-free 2D HOIP as superior room temperature spin generator.

RESULTS AND DISCUSSION

**Theoretical prediction of PEA-CuCl as an efficient spin generator via IRE**

Strong interfacial orbital hybridization brings about large IRE. Figure 1a shows the differential charge density diagram of PEA-CuCl/ferromagnet (FM) heterojunction (FM = Ni, Fe, Co; Ni is presented here as an example. See Methods and Figure S1-S2, Supporting Information for calculation details). In the diagram, yellow and cyan indicate electron accumulation and electron depletion, respectively. The accumulation and depletion of electrons at the interface interfacial between PEA-CuCl and Ni highlight substantial chemical charge transfer, consistent with Ni-Cl hybridization. This prediction is further experimentally verified by *in situ* X-ray photoemission spectroscopy depth profiling measurements (Figure S3, Supporting Information). Besides, the formation of Ni-Cl bonds can be clearly observed in the optimized CuCl-Ni supercell structure (Figure S1e), with bond length ranging from 2.194 Å to 2.473 Å, averaging of 2.347 Å (for comparison, the Ni-Cl bond length in $NiCl_2$ crystal is 2.426 Å[36]). The Ni-Cl bonding interaction also indicates the occurrence of orbital hybridization. As a result, strong orbital hybridization leads to the emergence of Rashba splitting bands near the Fermi level, suggesting the existence of large IRE (Figure 1b).

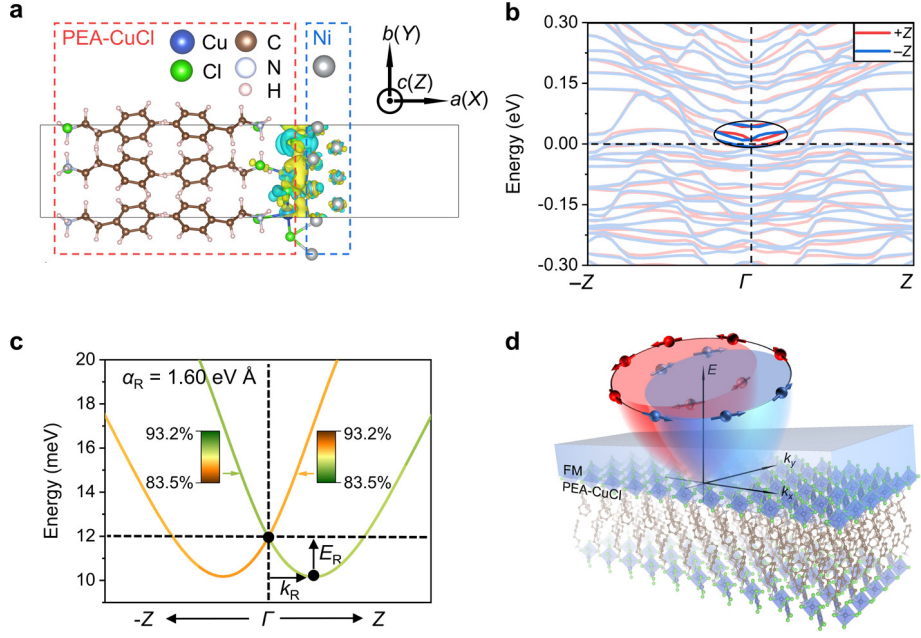

**Figure 1.** Prediction of PEA-CuCl/FM as an efficient spin generator *via* the IRE. (a) Calculated differential charge density with yellow and cyan indicating electron accumulation and electron depletion, respectively. (b) Band structure along $-Z$-$\Gamma$-$Z$ direction. Superimposed band structure with magnetization along $+Z$ (red line) and $-Z$ (blue line) are presented for clarity. (c) Orbital-resolved Rashba spin-splitting conduction bands of PEA-CuCl/Ni near the Fermi level, with color indicating the contribution of Ni $3d$ orbitals. Opposite color mappings are used for bands with magnetization along $+Z$ and $-Z$ for better visualization. (d) Schematic of spin-momentum locking due to IRE of PEA-CuCl/FM.

To further elucidate the connection between Rashba splitting bands (Figure 1b) and orbital hybridization, we present the orbital-resolved Rashba splitting bands near the Fermi level (10-20 meV) in Figure 1c. These Rashba splitting bands are mainly attributed to Ni $3d$ orbitals, and the rest are contributed from Cl $3p$ orbitals, confirming strong orbital hybridization. It is revealed by the color of the bands that represents the ratio of Ni 3$d$ orbitals, with opposite color mappings used for the two splitting bands to enhance visualization. The Rashba coefficient, $α_R$, is calculated to be as large as $α_R = 2E_R/k_R = 1.6$ eV Å. As a result, this large interfacial orbital hybridization is the key factor contributing to IRE of our lead-free PEA-CuCl, in contrast to the BRE of lead-based 2D HOIP, which arises from large SOC.

These Rashba splitting conduction bands of PEA-CuCl/Ni heterojunction endow it with charge-to-spin conversion ability. As depicted in Figure 1d, a Rashba effective field perpendicular to the electron wave vector $k$ can be expected. Consequently, when a charge current is introduced into PEA-CuCl/FM with the IRE, transversely polarized spin will accumulate at the interface to exert spin-orbit torques (SOTs)[4, 6] on the adjacent FM layer. The theoretical analysis above predicts that the lead-free 2D HOIP PEA-CuCl/FM can be an efficient spin generator.

**Experimental demonstration of IRE in PEA-CuCl for efficient spin generation**

Next, we adopt high-frequency electric transport measurements to experimentally investigate the IRE of PEA-CuCl/FM for spin generation. This method quantifies the Rashba effective field $h_{Rashba}$ *via* measuring its SOTs on the ferromagnetic resonance of adjacent FM, named as spin-torque ferromagnetic resonance (ST-FMR).[30, 37-41] ST-FMR has been widely used to investigate the IRE of many classical polycrystalline or amorphous all inorganic heterojunctions.[42-44] Compared with traditional indirect methods of characterizing the BRE in 2D HOIP, such as transient spectroscopies (Table S1, Supporting Information), the ST-FMR technique is more straightforward and powerful. It is because ST-FMR directly measures the SOTs generated by $h_{Rashba}$, which provide clear evidence for the charge-to-spin conversion process that is essential for spintronic devices.

The mechanism of ST-FMR is illustrated in Figure 2a (Methods, Supporting Information). A microwave current is introduced into the sample stripe through a bias tee, in conjunction with a magnetic field *H*. The IRE at the interface between PEA-CuCl and FM transforms the microwave current into an alternating $h_{Rashba}$. This alternating $h_{Rashba}$ produces SOTs to drive the magnetic moments of the adjacent FM layer into FMR, generating a DC voltage *via* the spin rectification effect.[45] The DC voltage is then extracted by the bias tee and measured by a nanovoltmeter.

Experimentally, PEA-CuCl thin films were prepared by spin-coating hydrothermally synthesized PEA-CuCl single crystal onto quartz substrates with a strong (100) texture (Methods, Figure S4, Supporting Information), followed by growing Py as the FM layer. A low-power DC sputtering technique is adopted for growing Py to minimize possible damage to the interface bewteen the organic layer and the FM layer.[46-49] Py alloy, predominantly composed of Ni, was chosen for its superior resonance performance, enabling accurate quantification of the Rashba effective field. High-angle annular dark field scanning transmission electron microscopy proves the high interfacial quality of our PEA-CuCl/Py heterojunction (Figure S5, Supporting Information). X-ray diffraction (XRD) measurement on PEA-CuCl/Py after one day of air exposure shows negligible degradation as compared with the as-grown sample (Figure S6, Supporting Information), ensuring the stability necessary to complete all ST-FMR measurements within a few hours. PEA-CuCl/Py stripes were then patterned and connected with thermally evaporated ground-signal-ground (G-S-G) coplanar waveguides (CPW) of Ti/Au to conduct microwaves for ST-FMR measurements.

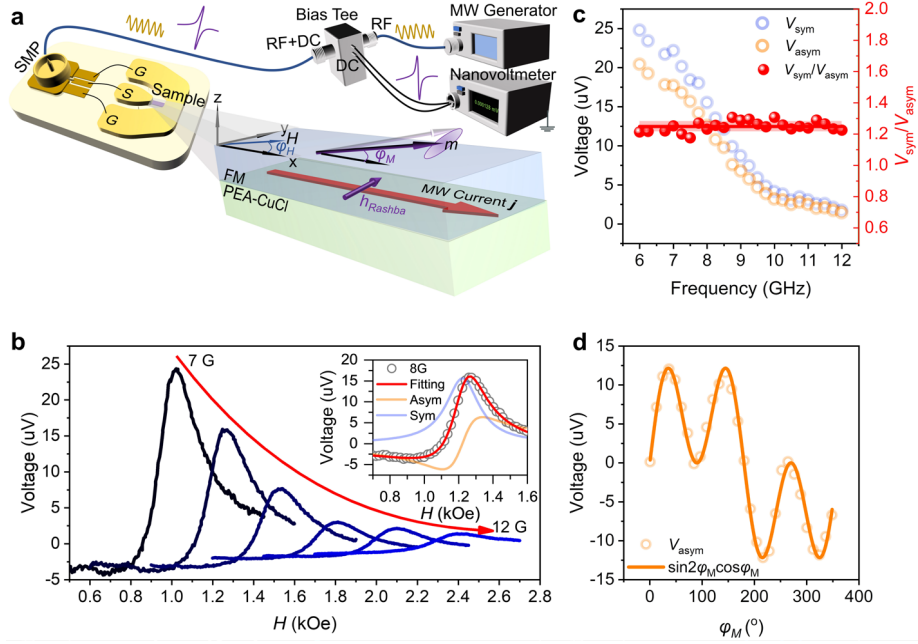

**Figure 2.** ST-FMR measurements for PEA-CuCl/Py. (a) Experimental set-up and working principles of ST-FMR. (b) *V-H* curves measured under 17 dBm microwave with frequency ranging from 7 GHz to 12 GHz at $\varphi_H = 45°$. The inset presents the raw *V-H* curve collected at 8 GHz as gray circles. The red line is the fitting result of the raw *V-H* curve, which is composed of both symmetric (blue line) and antisymmetric (yellow line) Lorentz lineshape. $V_{sym}$, $V_{asym}$, $\Delta H$, and $H_r$ are fitted to be 12.60 μV, 15.52 μV, 0.12 kOe, and 1.22 kOe, respectively. (c) Variation of $V_{sym}$, $V_{asym}$, and $V_{sym}/V_{asym}$ with microwave frequency *f*, denoted by blue, yellow, and red circles, respectively. The red line is the averaged value of $V_{sym}/V_{asym}$ at different frequencies, with the red shaded area representing the variance of $V_{sym}/V_{asym}$. (d) Angle-dependent $V_{asym}$ measured at 8 GHz, 17 dBm, and the corresponding fit using a $\sin2\varphi_M\cos\varphi_M$ term. $V_{asym}$ is fitted to be 15.80 μV, consistent with Figure 2b and 2c.

Figure 2b shows typical voltage traces when sweeping $H$ ($V$-$H$ curves) along an azimuthal angle $\varphi_H = 45°$ for PEA-CuCl/Py stripes, under excitation by 17 dBm microwaves at varying frequencies $f$. Due to the negligible magnetic anistropy, the azimuthal angle of $H$ ($\varphi_H$) is approximately equal to the azimuthal angle of magnetization ($\varphi_M$) (Note 1, Supporting Information). These $V$-$H$ curves exhibit both antisymmetric and symmetric Lorentz resonant lineshapes (inset of Figure 2b), indicating the exientence of SOTs on Py due to IRE. The resonant field $H_r$, linewidth $\Delta H$, symmetric voltage $V_{sym}$, and antisymmetric voltage $V_{asym}$ of $V$-$H$ curves under different $f$ can be fitted by $V = V_{asym}L_{asym} + V_{sym}L_{sym} = V_{asym}\Delta H(H - H_r)/[(H - H_r)^2 + \Delta H^2] + V_{sym}\Delta H^2/[(H - H_r)^2 + \Delta H^2]$. For higher $f$, a higher $H_r$ is required to meet the resonant Kittel formula $f = \frac{\gamma}{2\pi}\sqrt{H(H + 4\pi M_{eff})}$, with the gyromagnetic ratio $\gamma$ and the effective magnetization $4\pi M_{eff}$ for the Py layer fitted to be 181.8 GHz/T and 0.501 T, respectively (Figure S7, Supporting Information). The relatively small effective magnetization indicates a large perpendicular anisotropy $H_K$, which is consistent with a large interfacial Rashba spin-orbit coupling.[50, 51]

The IRE can also be verified by examining the relationship bewteen $f$ and $V_{sym}/V_{asym}$. As shown in Figure 2c, although both $V_{asym}$ and $V_{sym}$ decrease with increasing $f$ due to the reduced amplitude of the microwave current at higher frequency (Figure S8, Supporting Information), $V_{sym}/V_{asym}$ remains unchanged. This is a typical feature of ST-FMR signals arising from the IRE (Note 1, Supporting Information).[39] For comparison, in case of the spin Hall effect, which can also contribute to resonant $V$-$H$ curves, $V_{sym}/V_{asym}$ increases with increasing $f$ (see the ST-FMR results of Pt/Py for comparison in Figure S9, Supporting Information).[37]

We move on to demonstate that the IRE is the only origin of ST-FMR signals here. In principle, ST-FMR signals can originate from various effective fields in addition to the Rashba effective field $h_{Rashba}$, such as the Oersted field $h_{Oersted}$ and the Dresshaus spin-orbit field $h_{Dresshaus}$. When quantitatively deriving $h_{Rashba}$ from the ST-FMR signals, it is essential to distinguish these fields carefully.[30, 38, 39] Generally speaking, $V_{sym}$ and $V_{asym}$ are proportional to $\sin2\varphi_M h_z$ and $\sin2\varphi_M(h_x sin\varphi_M + h_y cos\varphi_M)$, respectively (Note 1, Supporting Information), if all possible components of the effective field ($h_x$, $h_y$, $h_z$) are considered. In this work, we focus on $V_{asym}$, as $V_{sym}$ may be due to the out-of-plane spin-orbit fields generated by the combined effects of in-plane spin polarization and exchange coupling, which is not related to in-plane $h_{Rashba}$.[30, 39, 52, 53] For $V_{asym}$, $h_x$ may originate from $h_{Dresshaus}$ while $h_y$ may be rooted in both $h_{Rashba}$ and $h_{Oersted}$. Thus, we performed angle-dependent measurements to distinguish them.

Figure 2d presents the angle-dependent $V_{asym}$ extracted from the $V$-$H$ curves measured at each $\varphi_M$, which can be well fitted by considering only the $h_y$-dependent $\sin2\varphi_M\cos\varphi_M$ term. We also carried out angle-dependent measurements for higher frequencies to cross-check, and $V_{asym}$ was consistently fitted by the $\sin2\varphi_M\cos\varphi_M$ term (Figure S10, Supporting Information). Hence, $h_{Dresshaus}$ along the $x$ axis does not exist in our PEA-CuCl/Py. Moreover, no significant resonance peak was found for a single Py layer on quartz (Figure S11, Supporting Information), excluding the Oersted field arising from the possible different conductivities of pre-grown and post-grown Py, consistent with former works.[37, 39, 40] Therefore, both $h_{Dresshaus}$ and $h_{Oersted}$ are carefully ruled out, leaving $h_{Rashba}$ as the sole contributor to $V_{asym}$ of the ST-FMR signal, which can be safely used to quantify the magnitude of $h_{Rashba}$.

For ST-FMR process stimulated by $h_{Rashba}$, $V_{asym}$ can be expressed as follows (See Note 1, Supporting Information for detailed derivations):

$$V_{asym}L_{asym} = -\frac{\Delta\rho IL}{2M_s}Re(\chi^{IP})h_{Rashba}\cos\varphi_M \sin2\varphi_M$$

Where $\Delta\rho$ is the magnitude of anisotropic magnetoresistance, $I$ is the amplitude of microwave current density $j$, $L$ is the length of the stripe, $M_s$ is the saturation magnetization of Py, and $Re(\chi^{IP})$ is the real component of the matrix element $\chi^{IP}$ of the complex dynamical magnetic susceptibility $\chi$. With $\Delta\rho = 0.155\%$, $I = 7\times10^9$ A m$^{-2}$, $L = 212$ μm, $M_s = 648$ emu/cc, and $\varphi_M = 45°$, $h_{Rashba}$ is calculated to be 0.98 Oe by fitting the $V_{asym}L_{asym}$ curve shown in Figure 3a, and $h_{Rashba}$ per unit current density is calculated to be 14.04 Oe per $10^{11}$ A/m$^2$ for 17 dBm microwave excitation of PEA-CuCl/Py.

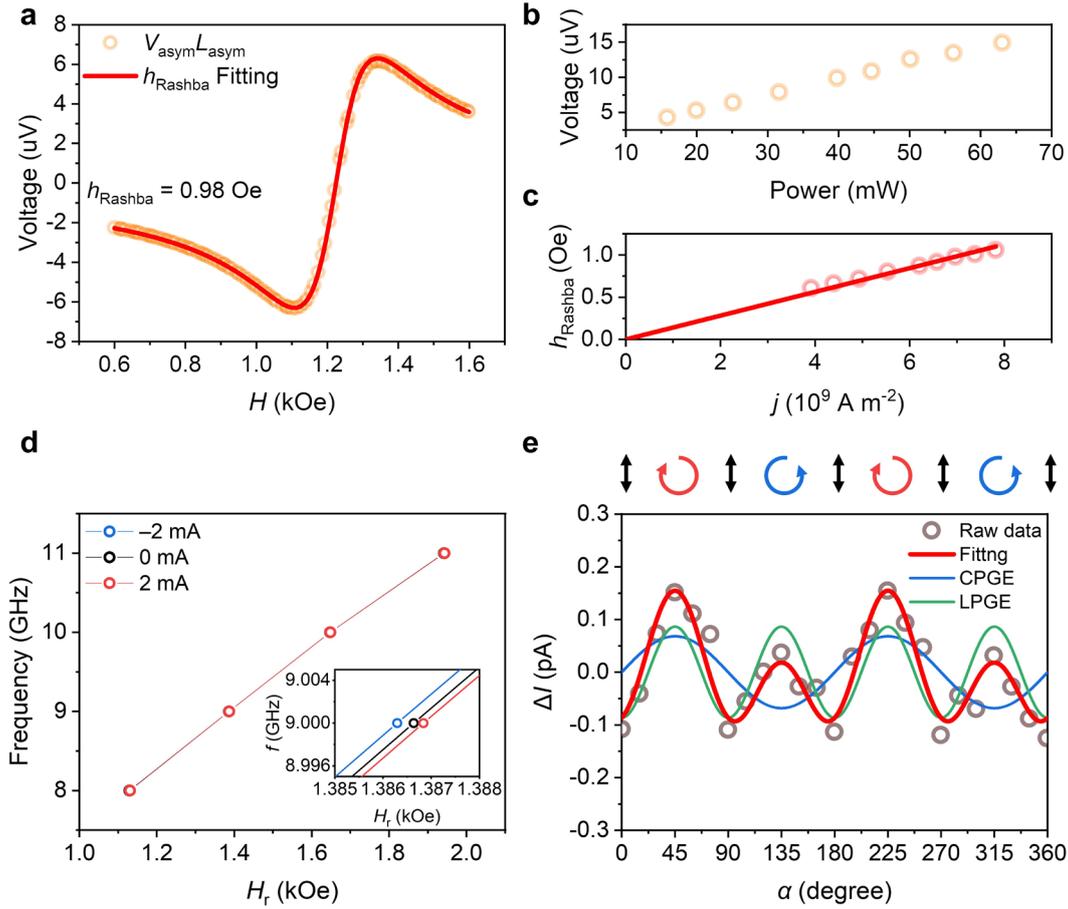

**Figure 3.** Demonstration of the IRE and its magnitude in PEA-CuCl/Py. (a) Fitted $V_{asym}L_{asym}$ of the *V-H* curve in the inset of Figure 2b, where corresponding $h_{Rashba}$ is fitted to be 0.98 Oe. (b) Linear growth of $V_{asym}$ with microwave power. (c) Linear relationship between the microwave current density *j* and the Rashba effective field $h_{Rashba}$. (d) Dispersion between microwave frequency and resonant field under DC current of 0 mA and ±2 mA. The inset shows a magnified view to highlight the shift in resonant fields. (e) Dependence of the room-temperature photogalvanic current perpendicular to the plane of incidence with the λ/4 plate rotation angle *α*, under the excitation of a 405 nm laser at an incident angle of 20°. CPGE: the circular photogalvanic effect, which is spin-dependent. LPGE: the linear photogalvanic effect, which is spin-independent.

We further performed several experiments to validate the IRE as the origin of $V_{asym}$. Figure 3b presents the power-dependent $V_{asym}$ obtained at $\varphi_M = 45°$, 8 GHz. $V_{asym}$ grows linearly with microwave power $P$, confirming that it remains in the linear response regime of excitation. By applying the same fitting process as in Figure 3a and calibrating the microwave current density $j$ at each $P$ (Figure S8, Supporting Information), a clear linear relationship between $h_{Rashba}$ and $j$ is established (Figure 3c). This linear relationship rules out possible thermal artifacts, which are quadratically dependent on $j$, indicating that $h_{Rashba}$ intrinsically originates from the IRE. Moreover, Py and PEA-CuCl thickness-dependent ST-FMR measurements are also consistent with the interfacial origin of the Rashba effect (Figure S12, Supporting Information).

To consolidate the conclusion, the existence of IRE has also been independently checked by DC-tuned ST-FMR measurements[37, 54-58] (Figure 3d, Figure S13, Supporting Information) and the circular photogalvanic effect[17, 19] (Figure 3e, Figure S14, Supporting Information). Specifically speaking, DC current was applied into PEA-CuC/Py stripe to generate additional $h_{Rashba}$, and the resulting ST-FMR spectra were collected by a lock-in amplifier. The resonant fields were extracted and summarized in Figure 3d. It is clear from the magnified inset that, the resonant field shifts in opposite directions for DC current of opposite polarities, further verifying the presence of $h_{Rashba}$ induced by IRE. Moreover, Figure 3e shows a distinct difference in the magnitude of photocurrent under right and left circularly polarized photoexcitation, characteristic of the circular photogalvanic effect, which indicates the Rashba spin-splitting band. This separately confirms the existence of IRE.

**Comparison of IRE in PEA-CuCl with control samples and benchmark systems**

We inserted a 2 nm Cu layer between the PEA-CuCl layer and the Py layer as a control sample and conducted similar ST-FMR measurements. As shown in Figure 4a, the ST-FMR signal for PEA-CuCl/Cu/Py drastically decreases as compared with that of PEA-CuCl/Py. That is because the direct contact and high-quality interface between PEA-CuCl and Py is eliminated, which blocks the orbital hybridization and results in a smaller $h_{Rashba}$. Note that the conductive Cu layer also generates an Oersted field in Py, leading to the measured $h_{effective}$ being a sum of $h_{Rashba}$ and $h_{Oe}$ (Figure S15, Supporting Information). The $h_{effective}/j$ of PEA-CuCl/Cu/Py is calculated to be 4.52 Oe per $10^{11}$ A m$^{-2}$ (Figure 4b), indicating that $h_{Rashba}$ is more than three times smaller than that of PEA-CuCl/Py. When PEA-CuCl is removed, Cu/Py exhibits negligible spin generation ability (Figure S11, Supporting Information). These control experiments prove that the interface of PEA-CuCl/Py is the key for the IRE.

Furthermore, a lead-based 2D HOIP, single crystal of $(C_6H_5CH_2CH_2NH_3)_2PbBr_4$ (PEA-PbBr) was synthesized (Methods) and ST-FMR measurements on PEA-PbBr/Py thin films were performed. This control sample is selected to investigate whether SOC or the orbital hybridization between Ni and Cl determines IRE. In PEA-PbBr/Py, both Pb and Br have larger atomic SOC than Cu and Cl, resulting a much stronger SOC effect than PEA-CuCl/Py, without the orbital hybridization between Ni and Cl. $h_{Rashba}$ per unit current density of 1.47 Oe per $10^{11}$ A m$^{-2}$ was quantified for PEA-PbBr/Py, an order of magnitude smaller than that of PEA-CuCl/Py (Figure 4b, Figure S16, Supporting Information). This can be well comprehended by the results of DFT calculations (Figure S17, Supporting Information), where interfacial chemical charge transfer is obviously weaker for PEA-PbBr/Py as compared with PEA-CuCl/Py. Compared with Cl, the electronegativity of Br is weaker, the atomic radius is larger, and thus it is more difficult to form bonds with Ni at the interface. Hence, the Ni-Br bond is longer than the Ni-Cl bond, indicating that there will be less overlap and hybridization between Br and Ni. As a result, the quench of interfacial orbital hybridization in PEA-PbBr/Py overwhelms the enhancement of atomic SOC, bringing about a much smaller IRE.

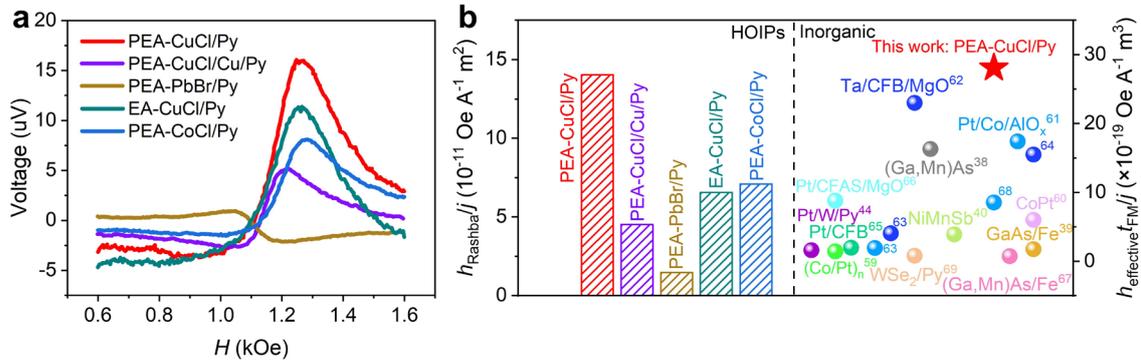

**Figure 4.** Comparison of IRE of lead-free PEA-CuCl/Py with other systems. (a) Comparison of $V$-$H$ curves for PEA-CuCl/Py with other HOIP systems measured at the microwave of 8 GHz, 17 dBm with $\varphi_H = 45°$. (b) Comparison of $h_{Rashba}/j$ and $h_{effective}t_{FM}/j$ for lead-free PEA-CuCl/Py with other 2D HOIP and traditional all-inorganic single layer and heterojunctions. CFB denotes CoFeB and CFAS represents $Co_2FeAl_{0.5}Si_{0.5}$.

The above results reveal that the interfacial orbital hybridization between Ni and Cl is vital to induce large IRE. Meanwhile, the organic cation PEA as well as the species of metal Cu can also influence this interfacial orbital hybridization, which tunes the IRE. Indeed, as shown in Figure 4, for $(C_2H_5NH_3)_2CuCl_4$/Py (EA-CuCl/Py, Figure S18, Supporting Information) and $(C_6H_5CH_2CH_2NH_3)_2CoCl_4$/Py (PEA-CoCl/Py, Figure S19, Supporting Information), the IRE is smaller than that of PEA-CuCl/Py, indicating that both Cu and PEA have a positive role on the large orbital hybridization between Cl and Ni.

Finally, we compare the Rashba effect in PEA-CuCl/Py with traditional all-inorganic single layer[38, 40, 59, 60] and heterojunctions[39, 44, 61-69] where Rashba effect dominates, such as Pt/Co/AlO$_x$ and Ta/CoFeB/MgO with noble metals (Figure 4b). To account for the influence of different thicknesses of ferromagnets in different systems, $h_{effective}t_{FM}/j$ is adopted as a figure of merit, where $t_{FM}$ is the thickness of the ferromagnet.[39] It turns out that our lead-free PEA-CuCl/Py is among the best, demonstrating the advantage of our strategy to explore the IRE of lead-free 2D HOIP PEA-CuCl for highly efficient spin generation. We also used the dimensionless spin torque efficiency $\xi$ to assist in comparing our IRE with other benchmark systems (Table S2, Supporting Information), which gives consistent conclusions. Please note that due to the different physical mechanisms between the Rashba effect and the spin Hall effect[70], $\xi$ can indicate the strength of the Rashba effect but may not quantitatively describe it.

CONCLUSION

In conclusion, we demonstrate 2D HOIP PEA-CuCl as an efficient room-temperature spin generator *via* the IRE both theoretically and experimentally. The large IRE of 2D HOIP PEA-CuCl is due to large interfacial orbital hybridization, supported by theoretical calculations and XPS measurements. A large $h_{Rashab}$ per unit current density of 14.04 Oe per $10^{11}$ A m$^{-2}$ is further quantified experimentally by systematic frequency-dependent, angle-dependent, and power-dependent ST-FMR measurements. Series of control experiments further cross-check the IRE. Compared with lead-based 2D HOIP and traditional all-inorganic heterojunctions, lead-free PEA-CuCl is not only more efficient but also nontoxic to human beings and cheaper to obtain. Our results contribute to the development of future flexible spintronic devices based on lead-free 2D HOIP, serving as spin generators at room temperature with superior properties.

METHODS

**Reagents:** Phenethylamine (PEA, 99.5%) and ethyl acrylate (EA, 99.5%) were purchased from Energy Chemical. Copper oxide (CuO, 99%), lead oxide (PbO, 99.999%), and cobalt(II) chloride hexahydrate (CoCl$_2$•6H$_2$O, 99%) were purchased from Sigma-Aldrich. 48% aqueous hydrobromic acid (HBr) was purchased from Meryer. 36.46% aqueous hydrochloric acid (HCl) was purchased from Sinopharm Chemical Reagent. All chemicals were used as received unless otherwise indicated. All the samples were kept in a glove box with O$_2$ and H$_2$O < 0.01 parts per million to prevent degradation or oxidation.

**Synthesis of (PEA)$_2$CuCl$_4$, (PEA)$_2$PbBr$_4$, (EA)$_2$CuCl$_4$, and (PEA)$_2$CoCl$_4$ single crystals:** To synthesize (PEA)$_2$CuCl$_4$, (PEA)$_2$PbBr$_4$, (EA)$_2$CuCl$_4$, and (PEA)$_2$CoCl$_4$ HOIP single crystals, 1 mmol PEA (EA) and 0.5 mmol CuO (PbO, CoCl$_2$•6H$_2$O) were mixed with 3 mL of HCl (HBr) solution loaded into a glass vial. The mixtures were subsequently dissolved at 80 °C in an oil bath. The reaction took about 1 hour to produce (PEA)$_2$CuCl$_4$, (PEA)$_2$PbBr$_4$, (EA)$_2$CuCl$_4$, and (PEA)$_2$CoCl$_4$ HOIP. The solutions of HOIP were slowly cooled to room temperature, with crystals precipitated. These crystals were then vacuum-filtrated and washed with diethyl ether. Final products were dried at 60 °C in vacuum overnight.

**Preparation of (PEA)$_2$CuCl$_4$, (PEA)$_2$PbBr$_4$, (EA)$_2$CuCl$_4$, and (PEA)$_2$CoCl$_4$ films:** Quartz substrates were washed sequentially by acetone, ethanol, and deionized water in a 100 W sonicator for 10 min each, followed by ultraviolet-ozone treatment for 20 min and an argon plasma treatment for 3 min. Solutions of (PEA)$_2$CuCl$_4$ were prepared by dissolving the perovskite crystals in mixed solvent (volume ratio of DMF:DMSO = 1:1) at different concentrations (4%, 6%, 8%). (PEA)$_2$CuCl$_4$ thin films were then prepared by spin-coating the solutions onto quartz substrates under 4000 rpm for 30 s, and then annealing at 80~90 °C for 10 min for crystallization with strong (100) textures. (PEA)$_2$PbBr$_4$, **(EA)$_2$CuCl$_4$, and (PEA)$_2$CoCl$_4$** films were also fabricated by spin-coating (using DMF as solvent) and annealing at 100 °C for 10 min.

**Preparation of (PEA)$_2$CuCl$_4$/Py, (PEA)$_2$PbBr$_4$/Py, (PEA)$_2$CuCl$_4$/Cu/Py, (EA)$_2$CuCl$_4$/Py, and (PEA)$_2$CoCl$_4$/Py for ST-FMR measurements:** To measure the spin-torque ferromagnetic resonance signals of HOIP, Py(20 nm) or Cu(2 nm)/Py(20 nm) were carefully deposited onto these thin films by magnetron sputtering at room temperature with a base pressure of 1×10$^{-5}$ Pa. The DC sputtering power was kept to be as low as 20 W with a corresponding low deposition rate of 0.13 Å/s to minimize possible damage to the organic layer. These (PEA)$_2$CuCl$_4$/Py, (PEA)$_2$PbBr$_4$/Py, (PEA)$_2$CuCl$_4$/Cu/Py, **(EA)$_2$CuCl$_4$/Py, and (PEA)$_2$CoCl$_4$/Py** thin films were then patterned into strips by tweezers manually with ~100 μm width and ~400 μm length. Ti(10 nm)/Au(100 nm) coplanar waveguide (CPW) were thermally evaporated using a mechanical mask to conduct microwaves. The exact dimensions of the stripes were measured by an optical microscope Nikon ECLIPSE LV150NL. Any contact with solvents like water, ethanol, or acetone was strictly avoided throughout device processing to prevent chemical damage to HOIP.

**ST-FMR set-up and measurements:** As shown in Figure 2a, microwave from a solid microwave generator is introduced into the stripe through the RF port of the bias tee, concomitant with a static magnetic field $H$ along an azimuthal angle $\varphi_H$ from the stripe. Microwave current $j(\omega)$ at the interface between PEA-CuCl and FM along $x$ axis generates alternating $h_{Rashba}$ along $y$ axis through the IRE. Alternating $h_{Rashba}$ produces SOTs that drive the magnetic moments of the adjacent FM layer into FMR around an azimuthal angle $\varphi_M$ from the strip, at the same frequency $\omega$. Because of the anisotropic magnetoresistance of FM, the resonance of FM causes its resistance to oscillate as $R(\omega) = R_0 + R_1(\omega)$. As a result, the voltage across the stripe $V = j(\omega)R(\omega)$ becomes $V(2\omega) + V(\text{constant})$, which is known as the spin rectification effect. The spin-rectified $V(\text{constant})$ is extracted by the bias tee and recorded *via* a nanovoltmeter through the DC and GND ports when sweeping $H$ under various azimuthal angles $\varphi_H$ and frequencies. A commonly used thermal-equivalent method[71, 72] was adopted to calibrate the microwave current (Figure S8, Supporting Information). For DC-tuned ST-FMR measurements, a DC current was applied into the stripe through the DC and GND ports of bias tee, and the ST-FMR spectra was collected by a lock-in amplifier.

**Circular photogalvanic effect measurements:** A semiconductor laser diode (405 nm) combined with a set of linear and circular polarizers was used as the photoexcitation source. The device was illuminated by the laser beam through a viewing port, and electrical signals were recorded *via* a source-meter unit (Keysight B2912A). The Schottky contact between electrodes and PEA-CuCl does not influence the qualitative demonstration of IRE.

**Characterizations:** Four-terminal transport measurements were carried out in a Physical Property Measurement System (PPMS) to collect the anisotropic magnetoresistance of Py. $\theta$-$2\theta$ XRD was carried out on Rigaku SmartLab. The saturation magnetization of Py layers was measured by a vibrating sample magnetometer. Atomic force microscopy was used to determine the roughness of HOIP films. XPS spectra were collected on a Thermo Scientific K-Alpha. The cross-sectional HAADF-STEM images were collected on an FEI Titan 80-300 electron microscopy equipped with a monochromator unit, a probe spherical aberration corrector, a post-column energy filter system (Gatan Tridiem 865 ER) and a Gatan 2k slow-scan charge-coupled device system, operating at 300 kV. The STEM sample was prepared by standard Focused Ion Beam (FIB) method using Zeiss FIB.

**DFT calculations:** Density functional theory calculations were performed in Vienna ab initio Simulation Packages (VASP).[73] Projector-augmented wave (PAW) pseudopotentials with Perdew-Burke-Ernzerhof exchange-correlation functional were used.[74, 75] The PEA-CuCl/FM models were constructed with two types of electrically neutral interfaces (CuCl/FM and –NH$_3$Cl/FM) to exclude the electrostatic influence, as shown in Figure S1, Supporting Information. PEA-PbBr/Ni reference models were calculated for comparison. Monkhorst-Pack grids[76] in k-space were set to be 1×3×3 for structural optimization and 1×5×5 for static self-consistent calculations. The van der Waals correction was incorporated, and spin-orbit coupling (SOC) was also considered for the Rashba effect. For the calculation of the Rashba splitting band structure, the magnetization of the Ni layer was constrained in Z and –Z direction[77] (in-plane direction of the heterojunction, consistent with ST-FMR measurements under an in-plane magnetic field). The inner force and free energy convergence criteria were set to 0.03 eV/Å and 10$^{-5}$ eV respectively, and a plane-wave cutoff energy of 520 eV. The modeling and visualization of HOIP/FM systems were performed in VESTA, and the planar-average potential was obtained from LOCPOT file by VASPKIT.

ASSOCIATED CONTENT

Supporting Information. DFT calculations for Type-I and Type-II interface (Figure S1), as well as control samples of PEA-CuCl/Fe, PEA-CuCl/Co (Figure S2) and PEA-PbBr/Ni (Figure S17); *in situ* XPS depth profiling measurements (Figure S3), basic material characterizations (Figure S4-S6); magnetic constants measurements (Figure S7); calibration of the microwave current (Figure S8); control measurements for Pt/Py (Figure S9), quartz/Py (Figure S11), Cu/Py (Figure S11), PEA-CuCl/Cu/Py (Figure S15), PEA-PbBr/Py (Figure S16), EA-CuCl/Py (Figure S18), PEA-CoCl/Py (Figure

S19), and angle-dependent ST-FMR measurements for PEA-CuCl/Py (Figure S10); Py and PEA-CuCl thickness-dependent ST-FMR measurements (Figure S12); DC-tuned ST-FMR measurements (Figure S13), circular photogalvanic effect measurements (Figure S14); cartesian coordinate systems (Figure S20); angle-dependent resonance field (Figure S21). Comparison of common methods to determine the magnitude of Rashba effect in HOIPs (Table S1). Comparison of $\xi$ of PEA-CuCl/Py with benchmark systems (Table S2).

.

AUTHOR INFORMATION

Corresponding Author

Cheng Song. Key Laboratory of Advanced Materials (MOE), School of Materials Science and Engineering, Tsinghua University, Beijing 100084, China. Email: songcheng@mail.tsinghua.edu.cn

Ying Lu. Technological Institute of Materials & Energy Science (TIMES), Xi'an Key Laboratory of Advanced Photo-Electronics Materials and Energy Conversion Device, School of Electronic Information, Xijing University, Xi'an 710025, China. Email: sunshine_ly327@163.com

Feng Pan. Key Laboratory of Advanced Materials (MOE), School of Materials Science and Engineering, Tsinghua University, Beijing 100084, China. Email: panf@mail.tsinghua.edu.cn

Author Contributions

L. H. Q. W., and Y. L. prepared the samples and carried out transport measurements. Q. W. carried out calculations. S.T. performed optical measurements. This work was conceived, led, coordinated, and guided by C. S. and F. P. All the authors contributed to the writing of the manuscript. All authors have given approval to the final version of the manuscript. ‡These authors contributed equally.

Notes

The authors have no conflicts to disclose.

ACKNOWLEDGMENT

This work is supported by the National Key Research and Development Program of China (Grant No. 2021YFB3601301), the National Natural Science Foundation of China (Grant No. 52225106 and 12241404), the Fundamental Research Funds for the Central Universities, China (2023JBZY002 and 2022YJS109), the Natural Science Foundation of Beijing, Municipality (Grant No. JQ20010), and the Senior Talents Foundation of Xijing University (Grant No. XJ23B03).